\documentclass[useAMS,usenatbib]{mn2e}

\usepackage{graphicx}
\usepackage{times}
\usepackage{natbib}
\usepackage{amsmath}

\newcommand{\gtrsim}{\ga}
\newcommand{\lesssim}{\la}

\def\zsun{{\rm Z_\odot}}
\def\zcrit{{\rm Z_{crit}}}
\def\msun{{\rm M_\odot}}
\def\msunh{{\rm M_\odot/{\it h}}}

\def\Zsun{{\rm Z_\odot}}

\def\Mpch{{\rm Mpc/{\it h}}}

\def\Omegab{{\Omega_{0,\rm b}}}
\def\Omegam{{\Omega_{0,\rm m}}}
\def\Omegal{{\Omega_{0,\rm \Lambda}}}

\def\NHInorm{N_{\rm HI, 20}}
\def\NHI{ N_{\rm HI}}

\title[DLA metals at $z\simeq 7$]{
Simulating extremely metal-poor gas and DLA metal content at redshift $z\simeq 7$
}
\author[U.~Maio, et al.]{
  Umberto~Maio$^{1,2,3}$\thanks{
    E-mail: maio@oats.inaf.it;
    $\quad$ {\it Marie~Curie~Fellow}
  },
  Benedetta Ciardi$^{4}$,
  Volker M\"uller$^2$
  \\
  ${}^1$INAF -- Osservatorio Astronomico di Trieste, via G.~B. Tiepolo, 11, 34143 Trieste, Italy\\
  ${}^2$Leibniz-Institut f\"ur Astrophysik (AIP), An der Sternwarte, 16, 14482 Potsdam, Germany\\
  ${}^3$Max-Planck-Institut f\"ur extraterrestrische Physik, Giessenbachstra{\ss}e 1,  D-85748 Garching b. M\"unchen, Germany\\
  ${}^4$Max-Planck-Institut f\"ur Astrophysik, Karl-Schwarzschild-Stra{\ss}e 1, D-85748 Garching b. M\"unchen, Germany\\
}

\begin{document}

\date{(draft)}
\pagerange{\pageref{firstpage}--\pageref{lastpage}}\pubyear{0}
\maketitle
\label{firstpage}

\begin{abstract}
  We present the first theoretical study of metals in damped-Ly$\alpha$ (DLA) systems at redshift $z\simeq 7$.
The features of cold, primordial gas are studied by means of N-body, hydro, chemistry simulations, including atomic and molecular non-equilibrium chemistry, cooling, star formation for population III and population II-I regimes, stellar evolution, cosmic metal spreading according to proper yields (for He, C, O, Si, Fe, Mg, S, etc.) and lifetimes, and feedback effects.
Theoretical expectations are then compared to recently available constraints from DLA observations.
\\
We find that DLA galaxies at $z\simeq 7$ account for $\sim 10$ per cent of the whole galaxy population and for most of the metal-poor galaxies at these epochs.
About $7$ per cent of these DLA galaxies contain purely pristine material and $\sim 34$ per cent of them consist of very weakly polluted gas, being, therefore, suitable candidates as population~III sites. The remaining $\sim 59$ per cent are enriched above $ \sim 10^{-4}\zsun$.
Additionally, DLA candidates appear to have:
gas masses  $\lesssim 2\times 10^8~\rm\msun$;
very low star formation rate, $\sim 10^{-3} - 10^{-2}~\rm\msun/yr$ (significantly weaker than late-time counterparts);
mean molecular fractions covering a fairly wide range, $x_{mol}\sim 10^{-3}- 10^{-6}$;
typical metallicities $Z\lesssim 3\times 10^{-3}~\zsun$
and HI column densities $\NHI \gtrsim 3\times 10^{20}~\rm cm^{-2}$.
They present no or weak correlations between their gas mass and $Z$, $\NHI$, or $x_{mol}$;
a moderate correlation between $x_{mol}$ and $Z$, linked to the ongoing molecular-driven star formation and metal pollution processes;
a mild anti-correlation between $\NHI$ and $x_{mol}$, due to H depletion into molecules;
and a chemical content that is subject to environmental dependencies.
\end{abstract}



\section{Introduction}\label{Sect:introduction}


Recent observations of the spectral absorptions around quasar (QSO) ULAS~J1120+0641 at $z\simeq 7$ \cite[][]{Mortlock2011} -- the farthest known QSO so far -- revealed, for the first time, the presence of extremely metal-poor gas in probable damped Lyman-$\alpha$ (DLA) systems at early times \cite[][]{Simcoe2012}, when the age of the Universe was only $\sim~$0.8~Gyr.
Such high-redshift observations are difficult because of the sparseness of bright quasars (that would allow direct measures of DLA absorptions) and because of the low DLA detection probability 
\cite[][]{Font-Ribera2012}.
However, from data at lower redshifts \cite[see e.g.][for reviews]{WolfeGawiserProchaska2005}, DLAs are believed to be absorption features due to the neutral, cold ($\lesssim 10^4\,\rm K$) gas clouds whose dense material (with typical HI column densities $\NHI \gtrsim 2\times 10^{20}\,\rm cm^{-2}$) removes light coming from background sources \cite[e.g.][]{Wolfe1986, Foltz1986, Smith1986}.
These structures host random motions of atomic gas on a wide area and are not usually found in proximity of very luminous, actively star forming regions
\cite[][]{Font-Ribera2012}.
DLA systems trace the faint end of the galaxy luminosity function \cite[e.g.][]{Fynbo2008}, and, hence, might represent those forming galaxies which are just building up their mass and are dominated by neutral gas \cite[e.g.][]{HaehneltSteinmetzRauch1998, ChenLanzetta2003, Cui2005, Nagamine2007, Tescari2009, Yajima2012}.
When found along the line of sight to high-redshift QSOs, they can be precious probes of the early cosmic environments and of the thermal state of the primordial Universe \cite[][]{Mortlock2011, Bolton2011arXiv}.
\\
Estimated upper limits for the claimed DLAs at $z\simeq 7$ give metallicities\footnote{
  For a gaseous system with a given chemical composition its metallicity $Z$ is defined as the mass fraction of all the heavy elements with atomic number larger than the one of helium. As a reference, the solar atmosphere has got a metallicity, indicated by $\Zsun$, of about 2 per cent, i.e. $\Zsun \simeq 0.02$.
According to \cite{Asplund2009} this value could be as low as 0.0134, though.
}
$Z\lesssim 10^{-4}-10^{-3}\zsun$ (according to whether the gas is gravitationally bound or not), and HI column densities for the cold, neutral gas of $\NHI\sim 10^{20.45}-10^{21}\,\rm cm^{-2}$
\cite[][]{Simcoe2012}.
If confirmed by future observations, they could be a viable site of primordial, pristine (population~III) star formation, thanks to such low metal content.
Similarly low values for $Z$ had never been found in previously detected DLAs (at $z \lesssim 5$), since their metallicities were at least one to two orders of magnitude higher, typically in the range $\sim 10^{-1}-1~\zsun$ \cite[e.g.][]{Heinmueller2006, Prochaska2007, Ellison2011}.
\\
Finally, molecular fractions, $x_{mol}$, measured in DLAs at redshift $1.8\lesssim z\lesssim 3$ seem not to be strongly correlated with $\NHI$, and show mean values of $\sim 10^{-4}$, but can even drop below $\sim 10^{-6}$  \cite[][]{Petitjean2000, Ledoux2003, Noterdaeme2008}.
These values were directly obtained from VLT observations of H$_2$ Lyman band emission and not inferred from CO conversion factor \cite[whose dependencies on galaxy type and metallicity might lead to an error of a factor of $\sim 10$; e.g.][and references therein]{Boselli2002, Galametz2011}.
\\
In the present work we focus on primordial, cold objects with low temperatures and containing molecular-rich gas.
We aim at exploring the chemical features of these cold, neutral structures at redshift $z\simeq 7$, by using numerical, N-body, hydrodynamical simulations \cite[][]{Maio2010}, taking into account atomic and molecular chemistry, star formation for both population~III (popIII) and population~II-I (popII-I) regimes, feedback effects, stellar evolution and metal spreading according to yields and lifetimes as function of stellar mass.
We will statistically compare our results with the most recent available observational constraints for high-$z$ DLAs \cite[][]{Simcoe2012} and we will address how the molecular component, driving star formation, relates with HI and $Z$ content.
\\
In Section~\ref{Sect:simulations} we give details about the code, the numerical implementation, and the simulation used here;
in Section~\ref{Sect:results} and \ref{Sect:discussion} we present and discuss our results;
and in Section~\ref{Sect:conclusions} we summarize our conclusions.


\section{Simulations}\label{Sect:simulations}


We consider the standard-$\Lambda$CDM simulations by \cite{Maio2010}, assuming an expansion parameter of $100 h\, \rm km/s/Mpc$ with $h=0.7$, and cosmological density parameters $\Omegam=0.3$, $\Omegab=0.04$ and $\Omegal=0.7$ for matter, baryons and cosmological constant, respectively, in a box with a side of 10 comoving \Mpch.
The run was performed with a modified version of Gadget2 code \cite[][]{Springel2005} and has a gas resolution of $m_{gas} = 3\times 10^5~\msunh$, sampling the cosmic medium with $2\times 320^3$ gas and dark-matter particles.
This set-up allows us to accurately resolve the gas behavior well below kpc-scales at redshift $z\simeq 7$.
Besides gravity and hydrodynamics, the implementation includes non-equilibrium atomic and molecular evolution for e$^-$, H, H$^+$, H$^-$, He, He$^+$, He$^{++}$, H$_2$, H$_2^+$, D, D$^+$, HD and HeH$^+$ \cite[][]{Yoshida2004, Maio2007}.
Since molecule formation drives gas cooling and collapse at early times, following a consistent reaction network for these different species is important to address correctly chemical and thermal patterns of the first structure formation sites. 
During gas collapse, molecules lead star formation
\cite[][]{Springel2003,Hernquist2003}, 
stellar evolution
\cite[][]{Tornatore2007met},
and consequent mechanical and chemical feedback, by also affecting gas cooling capabilities with additional resonant and fine-structure lines
\cite[][]{Maio2007}
and the built up of a cosmic UV background \cite[][]{HaardtMadau1996}.
Metal spreading and abundance determination \cite[][]{Tornatore2007met} are accurately accounted for by tracking the yields of He, C, O, Si, Fe, S, Mg, etc. \cite[][]{vdHG1997, Thielemann2003, WW1995, Woosley2002,HW2002, HW2010}, and by following the proper lifetime of stars with different masses \cite[][]{PM1993}.
This process determines the transition\footnote{
  This has been theoretically proven true both on cosmological \cite[][]{Maio2010} and on galactic \cite[][]{Ballero2006,Brusadin2013arXiv} scales.
}
from the pristine popIII to the enriched popII-I regime when a critical threshold metallicity $\zcrit=10^{-4}\,\zsun$ is reached \cite[][]{BrommLoeb2003, Schneider2003}\footnote{
  Despite the uncertainties \cite[e.g.][suggest $\zcrit=10^{-3.5}\zsun$]{SantoroShull2006}, more precise determinations of $\zcrit$ are not very relevant to the cosmological pollution history and the consequent star formation regime, as shown in \cite{Maio2010, Maio2011}.
}.
The assumed stellar initial mass function (IMF) is a top-heavy power law with a slope of $-2.35$ in the mass range [100,~500]~$\msun$ for popIII formation sites, while it is Salpeter in the mass range [0.1,~100]~$\msun$ for popII-I ones.
These mass ranges are consistent with theoretical models \cite[][]{ZeldovicNovikov1971,Portinari1998,Fryer2001,HW2002,Meynet2006} and observational evidences \cite[][]{Figer1998, Figer2005} of $\gtrsim 100\,\rm \msun$ stars.
In the popIII case, existence of stable supermassive $\gtrsim 750-1000\,\msun$ stars seems quite unlikely, being strongly related to several particular physical conditions, as abundance composition, rotation, radiative losses, and stellar companions \cite[see e.g.][]{HW2002, Marigo2003}.
Anyway, given the steepness of the IMF, the resulting metal spreading, energy output and stellar evolution properties would not be much affected, because weighted for the extremely small stellar fraction in the high-mass tail \cite[see also][]{Maio2010}.
Furthermore, $\gtrsim 100\,\rm \msun$ stars usually have all comparably small life-times.
As a result, the precise popIII mass upper limit does not appear to be crucial to gas cosmological evolution and star formation history.
\\
For each particle gravitational forces are computed and for SPH particles the hydro-equations are solved, as well, considering both cooling and heating in the energy balance (involving atomic and molecular emissions due to excitations, recombinations, ionizations, Bremsstrahlung and Compton effect -- additional gas heating can come from SN feedback and UV background at different times). 
This allows us to follow gas density, entropy, pressure, internal energy and temperature, and to properly update the different chemical species according to the local thermal conditions.
\\
Basic properties (masses, positions, velocities, radii, star formation rates, temperatures, metallicities, molecular abundances, HI column densities\footnote{
  Mean HI column densities are estimated by integrating the gas density in each halo along the radial direction.
},
etc.) of all the cosmic structures at redshift $z\simeq 7$ are retrieved with the help of friend-of-friend (FoF) and SubFind algorithms \cite[][]{Dolag2009}.
The algorithms identify structures by assuming a linking length of $20$ per cent the mean inter-particle separation and looking for the most bound particle in each halo.
From the constituting particles it is then possible to obtain all the information relative to each halo.
Numerically resolved objects \cite[][]{BateBurkert1997, Maio2009} are required to have a minimum of 200 gas particles\footnote{
  This means that the minimum {\it total} number of particles in each object is $\sim 10^3$ particles, since dark-matter particles are usually much more numerous than gas particles, from a factor of a few up to ten.
},
leading to a catalogue of 1522 'well resolved' structures.
We stress that our selection allows us to be complete for haloes with $\sim 10^3$ particles, consistently with e.g. \cite{Trenti2010}.
Since the fundamental nature of DLA systems is not clear, yet, we rely on the simple observational constraints of their thermal state and define DLA galaxies as the cold, neutral structures with a mass-weighted temperature\footnote{
  Alternatively, DLAs can be simply defined as those galaxies with mean HI column density higher than $\sim 2\times 10^{20}\,\rm cm^{-2}$. As it will be clear later, these two definitions are roughly equivalent. We prefer the temperature-based one, because it easily helps exclude very hot material and it is less affected by resolution limitations for density estimates or by sub-grid modeling for the unresolved interstellar medium \cite[see more discussions in e.g.][]{Tescari2009, Nagamine2007}.
}
below $ \sim 10^4\,\rm K$, at which Ly-$\alpha$ cooling drops rapidly.
This also ensures that the possible DLAs are located far away from powerful star forming regions, as mentioned in the Introduction. Indeed, hotter objects are unlikely to be fair candidates, as their ongoing star formation activity, that is mainly responsible for heating up the hosted gas in primordial haloes, would be in conflict with the lack of associations between DLAs and luminous, active galaxies.
We find that 122 of the resolved objects are DLAs and constitute roughly 10 per cent of the whole sample.
For further details on implementation and studies of the features of high-redshift structures, we refer the interested reader to our previous works\footnote{
See:
\cite{Maio2006, Maio2007, Maio2009, Maio2010, Maio2011},
\cite{Maio2011sup},
\cite{MaioIannuzzi2011},
\cite{Maio2011cqg},
\cite{Campisi2011},
\cite{PM2012},
\cite{Maio2012, Maio2013},
\cite{Akila2012},
\cite{Dayal2012arXiv},
\cite{Salvaterra2013}
and references therein.
}.
We point out, though, that the box and resolution used allow us to carefully describe primordial-gas inhomogeneities and ionization structure \cite[][]{Akila2012}, and to properly follow small- and intermediate-mass objects, that are very common at those redshifts.
The corresponding luminosity function at $z\simeq 7$ in the [$-22$, $-18$] UV magnitude range results in agreement with observational constraints and is consistent with gamma-ray-burst-host properties at this epoch \cite[][]{Campisi2011, Salvaterra2013}.
Very rare, high-$\sigma$ over-densities, lying in the exponential tail of the cosmological mass function might be missing, but, given their paucity and their expectedly hot temperatures and large luminosities, they should not change significantly our conclusions on DLA statistics, mostly at such early times.


\begin{figure*}
\centering
\includegraphics[width=0.9\textwidth]{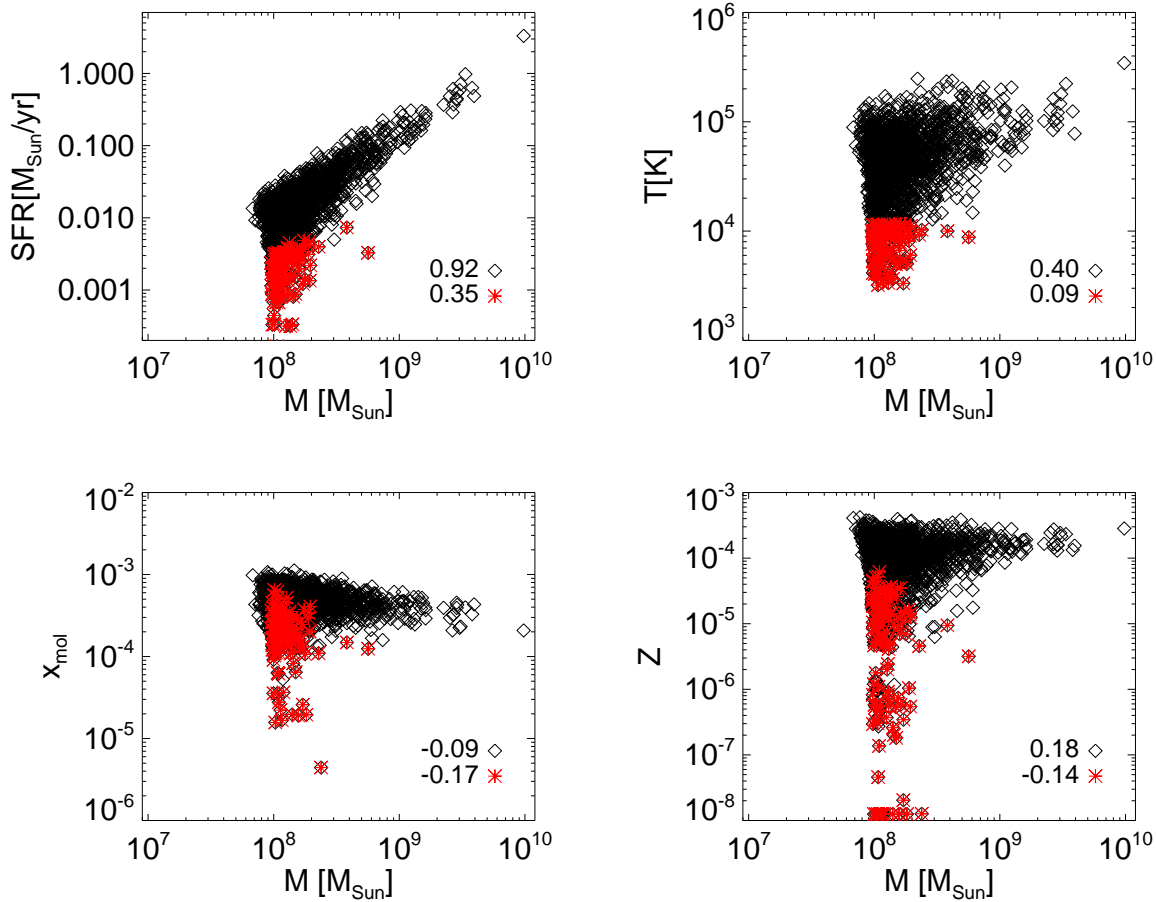}
\caption[]{\small
  Star formation rate, SFR (top left), mass-weighted temperature, $T$ (top right), mean molecular fraction, $x_{mol}$ (bottom left), and metallicity, $Z$ (bottom right), for the different objects as a function of their gas mass, $M$. In the latter plot metallicities with $Z<10^{-9}$ are added on the bottom axis for sake of completeness. Black diamonds refer to the whole sample, while red asterisks indicate DLA galaxies only (see text). The values on the bottom right corner of the panels refer to the correlation coefficient for each case.
}
\label{fig:all}
\end{figure*}


\section{Results}\label{Sect:results}


\noindent
In Fig.~\ref{fig:all}, we plot the trends for star formation rate (SFR), mass-weighted temperature, $T$, molecular abundance, $x_{mol}$, and metallicity, $Z$, as function of gas mass, $M$, for all the structures at $z\simeq 7$ (empty diamonds).
Additionally, DLA galaxies are over-plotted with asterisks.
In all cases the values of the corresponding correlation coefficients\footnote{
  The (Pearson) correlation coefficient of two stochastic variables is defined as the ratio between their covariance and the product of their standard deviations. A similar definition is adopted when considering the ranks of the data to compute the Spearman correlation coefficient. In the following, we will always refer to the Pearson correlation coefficient. We checked that the same conclusions are reached when looking at the Spearman one.
}
are quoted, as well.
\\
Typical gas masses at $z\sim 7$ range between several times $10^7\,\msun$ and $\gtrsim 10^{10}\,\msun$ and host star formation episodes whose rate increases roughly linearly with $M$, as expected from the \cite{Kennicutt1998}--\cite{Schmidt1959} law, from $\sim 10^{-2}~\rm\msun/yr$ up to $\sim 3~\rm\msun/yr$ (objects with null SFR are about 10 per cent of the DLA galaxies -- i.e. roughly 1 per cent of the whole sample).
\\
Our $z\simeq 7$ values for the DLA galaxies SFRs are $\lesssim 10^{-2}\,\rm\msun/yr$, significantly lower than those detected at more advanced cosmological times (estimates for DLA absorbers at $z\sim 2-3$ suggest typical $\rm SFR \sim 10^{-1}-1~\msun/yr$; e.g. \citealt{Dessauges2007,Heinmueller2006}), and lower-$z$ cosmological simulations usually find $\rm SFR \sim 10^{-1}-10~\rm \msun/yr$ \cite[e.g.][]{Tescari2009, Yajima2012, Cen2012}, while typical halo masses of $z\sim 2-3$ DLA galaxies range around $\rm\sim 6\times 10^{11}\msun$ \cite[][]{Font-Ribera2012}.
These discrepancies with $z\simeq 7$ objects are determined by the fact that primordial structures are generally smaller and contain less gas, hence their overall star formation activity is lower.
Smaller objects (with gas masses below $\sim 10^8\,\rm \msun$) are expected to have too shallow potential wells to hold hot gas and allow it to radiatively cool, so they would suffer gas photoevaporation processes by thermal and wind feedback, and could unlikely host significant DLA features \cite[][]{Nagamine2007, Tescari2009, Font-Ribera2012}.
\\
Simulated gas temperatures in our sample range from thousands to several $10^5\,\rm K$.
From a comparison with the previous panel, we see that objects hotter than $\sim 10^4\rm\, K$ have experienced star formation processes, which heated the ambient gas and ionized hydrogen.
Colder structures are usually more quiescent and can host neutral and/or molecular hydrogen.
\\
Mean molecular fractions are consistent with observational data\footnote{
Precise observational determinations of H$_2$ fractions are very complex
and have error bars which can be in excess of one dex.
}.
The distribution of $x_{mol}$ spans a few orders of magnitude and the resulting trend is a consequence of the local thermal conditions and of the interplay between gas cooling and star formation heating.
Larger mean molecular fractions ($x_{mol}\sim 10^{-3}$) are typical for smaller structures that contain significant HI gas \cite[as also expected in present-day galaxies;][]{Catinella2010} and that host H$_2$-cooling driven material which will soon ignite star formation.
Lower values can indicate either that gas densities are still low to boost molecule formation (this is likely for the low-mass end), or that heating due to stellar feedback has partially destroyed them, leaving only produced metals to sustain gas cooling and star formation (as  it is likely for the decreasing trend in the high-mass end).
In the cold, neutral DLA candidates the large scatter suggests that either molecules are still being formed by gas condensation, or that they have just been destroyed by environmental effects.
In fact, supernova explosions from newly born stars in the neighbouring regions can generate violent shocks (mechanical feedback) and cause rapid molecule dissociation with possible, partial re-formation behind them.
Furthermore, dynamical effects are responsible both for triggering molecule formation in compressed gas and, on the contrary, for stripping gas from smaller haloes.
At these epochs, a large reservoir of neutral atomic gas with $x_{mol} \ll 1$ is expected, though, since neutral H stays in a loitering, almost isothermal regime around $\sim 10^3-10^4\,\rm K$ for several $10^8\,\rm yr$ (comparable to the Hubble time at $z\sim 7$) before forming H$_2$ and collapsing into stars.
We underline that the plot refers to the mean H$_2$ fraction in each object, but locally $x_{mol}$ can vary dramatically of orders of magnitude due to the different thermal states in the interstellar medium.
These considerations suggest that the observed lack of a very tight correlation between H$_2$ and HI \cite[][]{Petitjean2000, Ledoux2003, Noterdaeme2008} is mainly due to the high sensitivity of molecules to the local environment, and we conclude that molecular content is not expected to evolve directly with redshift, but rather with the particular thermal evolution and star formation activity of the hosting structure (see next).
\\
Similarly, metallicities cover a large range of values, being quite broad at low masses and more tight at large masses, as a consequence of the different efficiency of metal spreading in different-mass objects.
DLA galaxies have typical metallicities below $10^{-4}$; $\sim 7$ per cent of them are completely devoid of metals, while $\sim 34$ per cent of them have got $Z <\zcrit$ (corresponding to $\sim 80$ per cent of all the haloes with sub-critical metallicity).
These sub-critical DLA galaxies represent $\sim 2.7$ per cent of the entire sample and contain almost completely pristine material, for they have not been significantly enriched by heavy elements, yet, and are ideal popIII star formation sites.
Differently from relatively low-redshift DLA metallicities
\cite[e.g.][]{Prochaska2007, Fynbo2010, Potzen2008, Tescari2009, Cen2012, Yajima2012},
$z\simeq7$ structures show much lower abundances, since the cosmic metal enrichment process in the Universe is still in its infancy.
At those early times ($< 0.8~\rm Gyr$), in fact, only a few massive stars could have exploded (as pair-instability and/or type~II supernovae) and polluted the primordial medium 
\cite[e.g.][]{TrentiShull2010, Maio2010, Maio2011}.
\\
Due to resolution limits, very small structures might be missing, but, they would be unlikely to host very dense gas with large molecular fractions, and would rather host material in the initial stages of the collapse process.
\\

\begin{figure*}
  \includegraphics[width=0.9\textwidth]{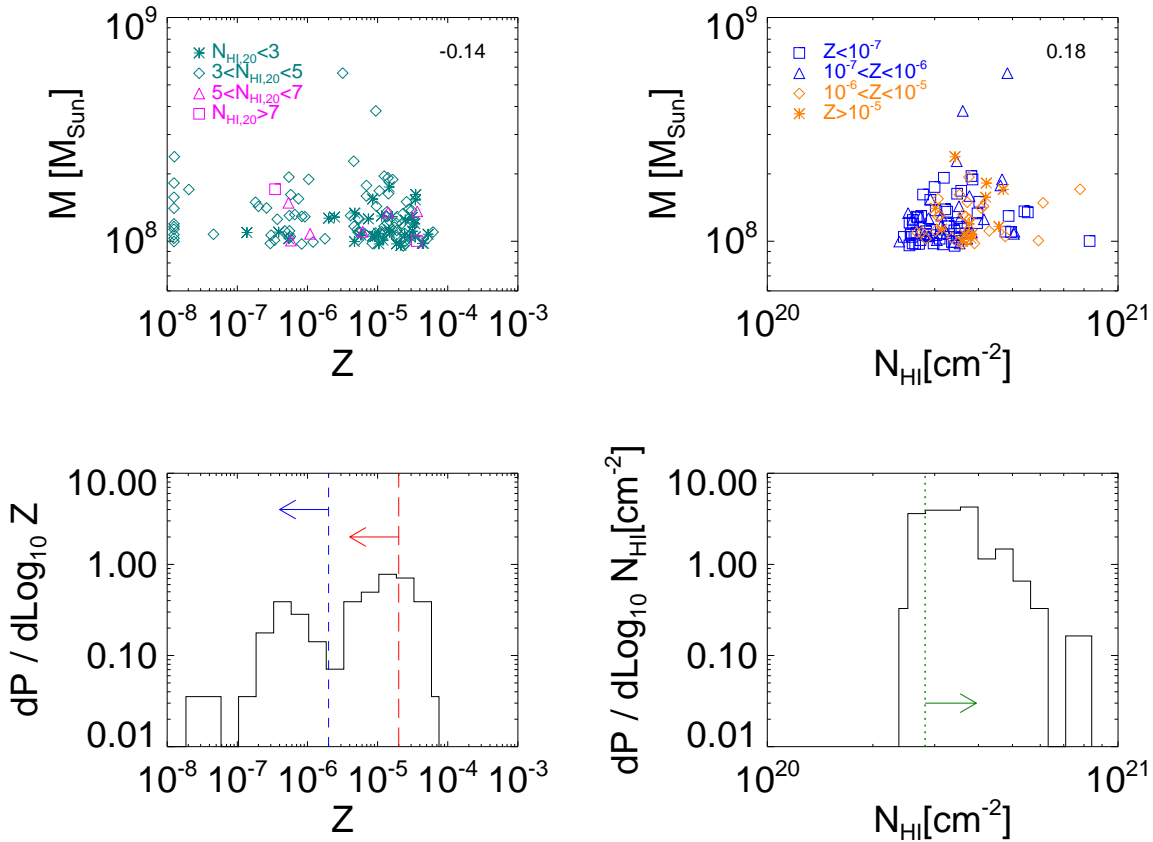}
  \caption[]{\small
    Scatter plot at $z\simeq 7$ for gas mass, $M$, and metallicity, $Z$ (top left); and for gas mass, $M$, and neutral-H column density, $\NHI$ (top right). For sake of completeness, data points for metallicity $< 10^{-9}$  are arbitrarily displayed on the left axis of the top-left panel. On the top-right corner of each panel the corresponding correlation coefficient is quoted. The differential probability distribution functions for $Z$ (bottom left) and $\NHI$ (bottom right) are displayed, as well. The vertical dashed lines are observational upper limits from \cite{Simcoe2012} and refer to gravitationally bound (long-dashed line at $Z=2\times 10^{-5}=10^{-3}\zsun$) and unbound (short-dashed line at $Z=2\times 10^{-6}=10^{-4}\zsun$) gas. The vertical dotted line is the lower limit for $\NHI$ at $10^{20.45}~\rm cm^{-2}$ \cite[][]{Simcoe2012}. The binnings of $Z$ (left) and $\NHI$ (right) distributions are base-10 logarithmic.
  }
  \label{fig:scatter}
\end{figure*}

\noindent
In Fig.~\ref{fig:scatter}, we focus on DLA galaxies only and display scatter plots for their main properties, namely, $\NHI$ and $Z$ versus the corresponding gas mass, $M$.
The $M - Z$ distribution is displayed for different $\NHI$ bins in units of $10^{20}~\rm cm^{-2}$, $\NHInorm$.
Similarly, the $M - \NHI$ distribution is plotted for different metallicity bins.
No clear relation with mass is visible since the correlation coefficients are close to zero.
When looking at the probability distribution functions and comparing them with suggested limits at $z\simeq 7$ \cite[][]{Simcoe2012}, we find a reasonable agreement between part of the simulated data and the observed upper limits.
We stress, though, that this could also be accounted for by the quite broad expected metallicity range, that covers a factor larger than $10^4$ in $Z$ (including even some pristine-gas objects) and peaks at relatively high $Z\sim 10^{-3}\zsun$.
\\
In addition, we remind that the local metal content of each object can vary significantly around the average values plotted in Fig.~\ref{fig:scatter}, due to the different cooling processes, star formation activity and feedback effects.
As a result, local metallicities in denser or more clustered gas will be much larger than in the peripheral outskirts of up to two dex within the same structure \cite[e.g.][]{Maio2010, Maio2011}.
\\
\begin{figure}
  \centering
  \includegraphics[width=0.45\textwidth]{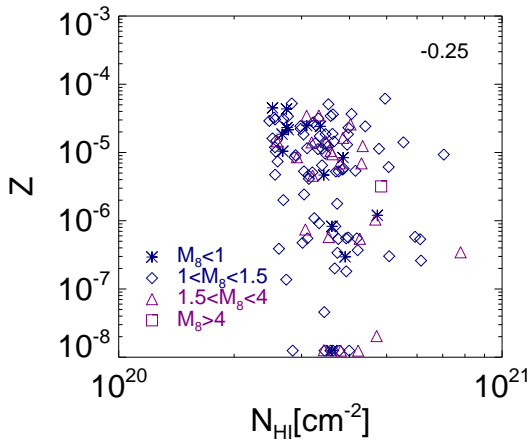}
  \caption[]{\small
    Scatter plot at $z\simeq 7$ for metallicity, $Z$, and neutral-H column density, $\NHI$. Different symbols refer to different masses (in units of $10^8\msun$, indicated with $M_8$). For sake of completeness, data points for $Z < 10^{-9}$ are added on the bottom axis. On the top-right corner the correlation coefficient is quoted.
  }
  \label{fig:scatterZNHI}
\end{figure}
Most of the simulated data present very low metallicities because they refer to cold structures that are usually located far away from hot star forming regions, the main sources of metal pollution.
The early spreading events have not reached them, yet, and the Universe contains still unpolluted areas \cite[e.g.][]{TrentiShull2010, Maio2010}.
In particular, the bimodality in the metal distribution with the dip around $2\times 10^{-6} \simeq 10^{-4} \zsun$ reflects the rapid transition from almost pristine popIII objects to highly-enriched popII-I objects\footnote{
Lower enrichment is due to the fact that these structures are isolated or more distant from actively polluting star forming sites.
We note that metal enrichment depends on the injection scheme and on the treatment of metal transport that plays a crucial role in the spreading process. The metal injection scheme is based on tracking SN explosions and AGB mass loss. The heavy elements produced are subject to transport via galactic winds with an assumed velocity of $\rm 500\,km/s$. A deeper discussion devoted to these issues can be found in \cite{Maio2010, Maio2011}.
}.
In fact, while in the low-metallicity end, objects are dominated by unpolluted or weakly polluted, popIII material, at larger $Z$ a popII-I regime is present.
The through highlights the rapid pollution process (due to the large metal yields of primordial stars) during the transition from one regime to another 
\cite[see detailed discussions in e.g.][]{Maio2010, Maio2011}.
\\

\noindent
Neither the trends for $Z$ and $\NHI$ are significantly correlated, as displayed in Fig.~\ref{fig:scatterZNHI}. This is consistent with lower-$z$ observations by e.g. 
\cite{Prochaska2003, Prochaska2007} and \cite{Ellison2012} and suggests a significant environmental dependence of chemical properties during structure formation.
\\

\begin{figure*}
\hspace{-0.8cm}
\includegraphics[width=0.9\textwidth]{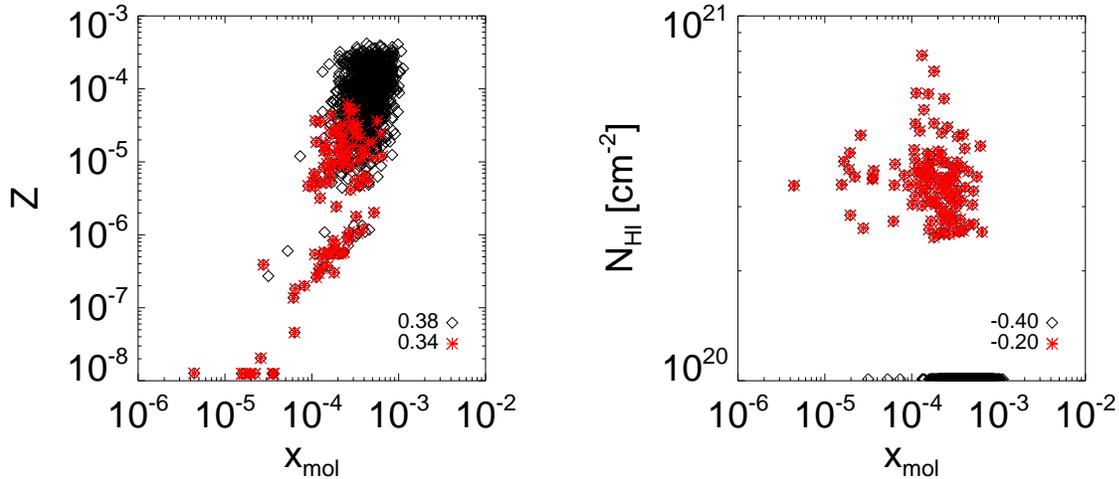}
\caption[]{\small
  Scatter plot at $z\simeq 7$ for metallicity, $Z$, versus molecular fraction, $x_{mol}$ (left), and neutral-H column density, $\NHI$, versus molecular fraction, $x_{mol}$ (right), for the whole sample (black diamonds) and DLA galaxies only (red asterisks). For sake of clarity, molecular fractions for $Z<10^{-9}$ and null $\NHI$ values are shown on the bottom of the left and right panel, respectively. The correlation coefficient is quoted on the bottom-right corner of each panel. 
}
\label{fig:scattermol}
\end{figure*}

\noindent
Finally, in Fig.~\ref{fig:scattermol} we display the trend of $Z$ and $\NHI$ with $x_{mol}$.
In the $x_{mol}$ versus $Z$ plot, we see a moderate tendency of finding metal-poor objects in correspondence of weak molecule formation. This is a sign of the little enrichment in quiescent, isolated, non-star-forming structures. Metal pollution becomes more important with the growth of stellar mass induced by molecular-driven gas collapse.
Hence the increasing behaviour of $Z$ with $x_{mol}$.
\\
Mean molecular fractions are generally too low to significantly affect $\NHI$, mostly for DLA objects.
However, the correlation coefficients suggest a slight trend of decreasing $\NHI$ for increasing $x_{mol}$ when looking at the whole sample, as a result of H ionization around $\sim 10^4\,\rm K$ and subsequent availability of free protons and electrons that can boost molecule formation via H$^+$ channel and deplete HI into H$_2$.
This anti-correlation is suggested also by the mildly negative correlation coefficients quoted in the legend.


\section{Discussion}\label{Sect:discussion}


Observational determinations of the chemical conditions of the primordial Universe at redshift $z\simeq 7$ have recently been claimed through studies of DLA HI column densities \cite[][]{Simcoe2012}.
Upper limits reported for the metallicity have been estimated around $\lesssim 10^{-4} - 10^{-3}\zsun$, supporting the existence of possible sites for popIII star formation.
As DLA galaxies are faint, rare objects with little star formation that are just building up their stellar mass, detections at high redshift are particularly challenging and thus theoretical works have mainly focused on later, $z\lesssim 4$, epochs \cite[e.g.][]{HaehneltSteinmetzRauch1998, Nagamine2007, Tescari2009, Cen2012, Yajima2012, Erkal2012}.
Early galaxies forming in H$_2$-cooling haloes need to be addressed by detailed molecular calculations, self-consistent stellar evolution and metal spreading according to proper yields and lifetimes for the different stellar population regimes.
\\
DLA galaxies are interesting tools to study dark ages, primordial cosmic environments, molecular content of high-$z$ objects, the regimes in which popIII stars could arise, metal spreading and the thermal state of the Universe during its first billion years.
\\
This is the first theoretical study of DLA metallicity at $z\simeq 7$, performed by looking at the physical and chemical features of primordial, cold, neutral structures and comparing them against recently available limits at $z \simeq 7$.
In the present work, we have assessed the basic properties of primordial galaxies at redshift $z\simeq 7$, by using detailed N-body, hydro, chemistry simulations \cite[][]{Maio2010}, including atomic and molecular network (for e$^-$, H, H$^+$, H$^-$, He, He$^+$, He$^{++}$, H$_2$, H$^+$, D, D$^+$, HD and HeH$^+$), chemical-dependent cooling, self-consistent stellar evolution for popIII and popII-I star formation regimes, metal spreading for individual species (C, O, Si, Fe, S, Mg, etc.) according to proper yields and stellar lifetimes, and feedback mechanisms.
\\
We stress that the theoretical success in quantifying atomic and molecular properties of cosmic structures at different times relies mainly on the correct determination of the features of their elementary constituents, in particular, gas chemical composition and stellar evolution in a cosmological context.
These are necessary ingredients to properly address the existence of metal-poor gas at redshift $ z\simeq 7 $.
\\
Generally, results are not strongly dependent on the precise values of cosmological parameters, but rather on the baryon physics within formed structures\footnote{
  Fluctuations in our findings are led by changes in hydrodynamics, cooling, star formation, stellar evolution and stellar properties, that are the key mechanisms during baryonic structure growth and have much more important implications than the underlying cosmological model, mostly at high $z$. We checked and discussed such issues in
\cite{Maio2006, Maio2009, Maio2010, MaioIannuzzi2011}.
}.
Hence, the whole statistical analysis would not show significant deviations and would be left roughly unchanged in alternative cosmologies.
\\
On the other hand, we expect slight variations in the theoretical results due to the different approximations about, e.g., SN fraction and energy, stellar evolution parameters, metal yields, wind energy, finite box size, resolution, etc. \cite[as extensively discussed in][]{Maio2010}.
In general, even changing the details of the IMF slope, the SN fraction does not change significantly for fixed IMF mass range and its variation is likely to have negligible impacts.
The values of SN energies are tightly bracketed around $\sim 10^{51}\,\rm erg$ and thus they cannot vary too much to affect our findings significantly.
Similarly, also stellar evolution parameters are quite well known \cite[see detailed studies in][and references therein]{Greggio2005}.
There is no wide consensus on metal yields, never the less, in a previous works \cite[see][]{Maio2010}, we tested a number of possible predictions and found a general agreement on the amounts of metals produced at early times.
Galactic winds represent a more complicated problem, however, several observations \cite[][]{Martin1999, Pettini2002, Bouche2012a, Newman2012a, Bouche2012b, Newman2012b} show that galactic outflow rates are usually a few times the star formation rate.
We assumed in this work a value of $\eta\sim2$ for the mass loading rate, hence we are confident on our results.
Larger $\eta$ would probably lead to lower $Z$, still in the direction suggested by \cite{Simcoe2012}.
\\
Numerical effects might play some role, because the finite size of the box inevitably determines an upper mass cut-off, as rarer bigger galaxies are not included.
Due to their paucity, though, this should not degrade our general conclusions.
Finally, we note that resolution artifacts are possibly minor issues: the kind of simulations that we have used in the present work usually shows convergence at $z \gtrsim 10$, after which star formation histories and various different physical quantities tend to follow similar trends \cite[see][ for further details]{Maio2010, Maio2011, MaioIannuzzi2011, Salvaterra2013}.


\section{Conclusions}\label{Sect:conclusions}


Thanks to a self-consistent inclusion of the relevant items, we are able to investigate theoretical properties of early metal-poor gas and to compare against recent observational constraints of claimed DLAs at $z\simeq 7$ \cite[][]{Simcoe2012}, such as $Z$, $\NHI$ and molecular fractions.
\\
We sample primordial galaxies with gas mass between several $10^7~\msun $ and $\sim10^{10}~\msun$, at $z\simeq 7$.
They are expected to have:
\begin{itemize}
\item
  typical $\rm SFR\sim 10^{-2}-3~\rm\msun/yr$, increasing roughly linearly with gas mass;
\item
  significant environmental dependency of molecular fractions, that can be
  boosted in dense star forming sites, but also rapidly destroyed by stellar
  feedback mechanisms;
\item
  broad range of metallicities, from $Z\sim 0$ to a few $10^{-2}\zsun$,
  since heavy elements are better trapped in larger potential wells, while
  they are easily lost in shallower potential wells;
\item
  dominant popII-I star formation regime.
\end{itemize}
Among such structures, $z\simeq 7$ DLA galaxies show:
\begin{itemize}
\item
  typical gas masses $\lesssim 2\times 10^8~\rm \msun$;
\item
  very low SFR,  $\sim 10^{-4}-10^{-2}~\rm \msun/yr$ (significantly weaker than lower-$z$ counterparts);
\item
  broad range of mean molecular fractions, $\sim 10^{-3}- 10^{-6}$;
\item
  typical $\NHI \sim 3\times 10^{20}~\rm cm^{-2}$;
\item
  typical mean metallicities $Z\lesssim 10^{-3}\zsun$, and the possibility of hosting popIII star formation;
\item
  weak correlations between gas mass, $M$, and typical chemical properties, as $Z$, $\NHI$, $x_{mol}$;
\item
  moderate correlation between $Z$ and $x_{mol}$, due to molecular-driven star formation and metal pollution;
\item
  moderate correlation between $\NHI$ and $x_{mol}$, due to H depletion into molecules;
\item
  evident environmental dependencies of chemical properties.
\end{itemize}
Simulated DLA galaxies represent the most of the pristine-gas-hosting galaxies.
In $\sim 34$ per cent of the cases they contain very metal-poor gas and in $\sim 7$ per cent of the cases they consist of completely unpolluted material.
Therefore, they are ideal site to look for popIII star formation at redshift $z\simeq 7$ and higher.\\
We finally highlight that the observational, preliminary constraints on metal-poor gas at $z\simeq 7$ give only weak upper limits for $Z$ and actual values could be much lower or closer to $\sim 0$, mostly in the case that the claimed DLAs would turn out to be just diffuse gas clumps (indeed, such very low metallicities would be in tension with the usually larger $Z$ that we expect for the majority of the cases).
This would be still interesting, though, because would be a more direct confirmation of the existence of pristine material at relatively advanced cosmic epochs (when PISN, SNII or AGB stars are expected to have already polluted the surrounding medium).
In this respect, further studies, observational confirmations and more detailed high-$z$ data would be desirable to investigate the enrichment history of the Universe and to unveil the role of different physical mechanisms in different cosmological environments.


\section*{acknowledgments}
We acknowledge the referee, Mike Shull, for his helpful and constructive comments that enabled us to improve the presentation of this work.
We also acknowledge useful discussions with J.~Bolton, S.~Borgani, E.~Tescari, M.~Viel.
U.~M.'s research leading to these results has received funding from a Marie Curie fellowship of the European Union Seventh Framework Programme (FP7/2007-2013) under grant agreement n. 267251.
The calculations were performed on the Hydra cluster at the computing center of the Max Planck Society, Rechenzentrum Garching b. M\"unchen (RZG).
For the bibliographic research we made use of the tools offered by the NASA Astrophysics Data System.


\bibliographystyle{mn2e}
\bibliography{bibl.bib}

\label{lastpage}
\end{document}